\documentclass[
superscriptaddress,
reprint,
 amsmath,amssymb,
 aps,
]{revtex4-2}

\usepackage{upgreek}
\usepackage{graphicx}
\graphicspath{{figures/}}
\usepackage{dcolumn}
\usepackage{bm}
\usepackage{hyperref}
\usepackage[mathlines]{lineno}
\usepackage{ulem}
\usepackage{multirow}
\usepackage{booktabs}
\usepackage{caption}
\usepackage{subcaption}
\usepackage{siunitx}
\usepackage{xcolor,colortbl}
\usepackage{array}
\usepackage{tabularx}
\usepackage{footmisc}
\usepackage{threeparttable} 

\begin{document}


\title{Precision measurements on Si$^-$}

\author{J. Karls}
\affiliation{%
 Department of Physics, University of Gothenburg, SE-412 96 Gothenburg, Sweden
}%
\author{H. Cederquist}
\affiliation{%
 Department of Physics, Stockholm University, AlbaNova, SE-106 91 Stockholm, Sweden
}%
\author{N. D. Gibson}
\affiliation{%
 Department of Physics and Astronomy, Denison University, Granville, Ohio 43023, USA
}%
\author{J. Grumer}
\affiliation{
Department of Physicsand Astronomy, Theoretical Astrophysics, Uppsala University, Box 516, SE-75120 Uppsala, Sweden
}%
\author{M. Ji}
\affiliation{%
 Department of Physics, Stockholm University, AlbaNova, SE-106 91 Stockholm, Sweden
}%
\author{I. Kardasch}
\affiliation{%
 Department of Physics, University of Gothenburg, SE-412 96 Gothenburg, Sweden
}%
\author{D. Leimbach}
\affiliation{%
 Department of Physics, University of Gothenburg, SE-412 96 Gothenburg, Sweden
}%
\author{P. Martini}
\affiliation{%
 Department of Physics, Stockholm University, AlbaNova, SE-106 91 Stockholm, Sweden
}%
\author{J. E. Navarro Navarrete}
\affiliation{%
 Department of Physics, Stockholm University, AlbaNova, SE-106 91 Stockholm, Sweden
}%
\author{R. Poulose}
\affiliation{%
 Department of Physics, Stockholm University, AlbaNova, SE-106 91 Stockholm, Sweden
}%
\author{S. Rosén}
\affiliation{%
 Department of Physics, Stockholm University, AlbaNova, SE-106 91 Stockholm, Sweden
}%
\author{H. T. Schmidt}
\affiliation{%
 Department of Physics, Stockholm University, AlbaNova, SE-106 91 Stockholm, Sweden
}%

\author{A. Simonsson}
\affiliation{%
 Department of Physics, Stockholm University, AlbaNova, SE-106 91 Stockholm, Sweden
}%
\author{H. Zettergren}
\affiliation{%
 Department of Physics, Stockholm University, AlbaNova, SE-106 91 Stockholm, Sweden
}%
\author{D. Hanstorp}
\affiliation{%
 Department of Physics, University of Gothenburg, SE-412 96 Gothenburg, Sweden
}%

\date{\today}

\begin{abstract}
High-precision measurements of the electron affinities (EA) of the three stable isotopes of silicon, $^{28}$Si, $^{29}$Si and $^{30}$Si, have been performed at the cryogenic electrostatic ion-beam storage ring DESIREE. The quantum states of the ions were manipulated using laser depletion, and the ions were photodetached by laser photodetachment threshold spectroscopy. These EA values are the first reported for  $^{29}$Si$^-$ and $^{30}$Si$^-$ and provide a reduced uncertainty for $^{28}$Si$^-$. The resulting EAs are $EA(^{28}$Si$) = 1.38952201(17)$ eV, $EA(^{29}$Si$) = 1.38952172(12)$ eV and $EA(^{29}$Si$) = 1.38952078(12)$ eV, with the corresponding isotope shifts $IS(^{29-28}$Si$) = 0.29(16) \upmu$eV and  $IS(^{30-28}$Si$) = 1.23(16) \upmu$eV. In addition to these measurements, the resolution and signal-to-background level was sufficient to reveal the  hyperfine structure splitting in the $^{29}$Si$^-$ isotope, which we report to be $1.8(4)\upmu$eV. 

\end{abstract}

\maketitle


\section{\label{sec:introduction}Introduction}
Atomic negative ions are not as well mapped out as their neutral and positive counterparts. Negative ions do not experience a long range Coulomb potential, but instead the electron correlation becomes relatively more important. This make studies of negative ions an excellent approach to benchmarking theoretical models that goes beyond the independent particle model. 

Today, most elements that can form a negative ion have had their electron affinities (EA) experimentally determined, with the exception of most elements that do not form any stable isotopes. However, the EAs for some radioisotopes have also been investigated \cite{Rothe2017Laser128I-_2,Leimbach2020TheAstatine}, and the search for the EAs of all elements in the periodic table is well underway. 

Due to the spatial extension and finite mass of the nucleus, different isotopes of the same elements will have slightly different EAs. Isotope shifts in the EA have been measured for the negative ion isotopes of $^{2-1}$H, \cite{Lykke1991ThresholdH-_2}, $^{13-12}$C  \cite{Bresteau2016IsotopeMicroscopy}, $^{17-16}$O \cite{Blondel2001Electron17O-_2}, $^{18-16}$O,  \cite{Kristiansson2022PrecisionIons} and $^{37-35}$Cl \cite{Berzinsh1994IsotopeChlorine}.
The isotope shift (IS) consists of three parts; the normal mass shift (NMS), the specific mass shift (SMS) and the field shift (FS): 
\begin{equation}
    \ IS = NMS + SMS + FS.
\end{equation}

The NMS is trivial to calculate since one simply replaces the electron mass with its reduced mass. The SMS is caused by electron correlation where a correlated/uncorrelated motion increase/decrease the isotope shift. This is a pure atomic physics property. The FS, on the contrary,  is a pure nuclear physics effect where the spatial extension of the nucleus causes a small shift in the atomic potential. The FS is an established method to study nuclear properties to obtain nuclear information such as nuclear deformation \cite{Koszorus2021Charge32} or  nuclear moments \cite{deGroote2017Dipole/math_2}. However, the only measureable quantity is the total isotope shift (IS).
Hence, nuclear physicists need accurate atomic models to calculate the mass shift contributions in order to extract the FS.  

In lighter elements, the mass shift dominates. This means that investigations of isotope shifts on light elements can be used to develop precision atomic theory that can predict the specific mass shift. Here the IS in the EA is the ideal test ground since both the EA itself and the SMS are very sensitive to electron correlations. Such studies are hence both of interest within atomic physics to probe electron correlation effects and as a tool critically needed in nuclear physics. 

In addition to the IS, the hyperfine structure (HFS) in negative ions has been studied experimentally for $^{17-16}$O\cite{Trainham1989Measurement33S-} where a radiofrequency method was used to induce transitions between the hyperfine levels, and in Os$^-$, where the HFS was resolved in a bound-bound transition \cite{Fischer2010FirstAnion}. Further, the HFS of the Si$^-$ homologue $^{17}$O$^-$ has been investigated theoretically \cite{Blondel2001Electron17O-_2}.  

Negative ions can be studied either in an ion trap, which yields long interaction times, or in an ion beam apparatus, which offer high resolution due the Doppler compression and efficient detection of fast neutralized atoms. In a storage ring these advantages are combined. Here, ion beams can be stored for an extended period of time, which give several experimental advantages when studying negative ions. Cryogenic electrostatic ion-beam storage rings have developed quickly in the recent years \cite{Thomas2011TheStudies,Schmidt2013FirstDESIREE,vonHahn2016TheCSR,Nakano2017DesignRICE}, allowing for measurements of both structural and dynamical  properties of negative ions \cite{Backstrom2015Storing-_2,Kristiansson2022High-precisionOxygen,Mull2021MetastableRing_2,Kristiansson2021ExperimentalIr-_2,Kristiansson2022MeasurementBi-_2}. The Double ElectroStatic Ion-Ring ExpEriment (DESIREE) is a cryogenic double storage ring facility, where ion beams can be stored for thousands of seconds \cite{Thomas2011TheStudies,Schmidt2013FirstDESIREE,Backstrom2015Storing-_2,Schmidt2017RotationallyDESIREE_2}.

Experimental studies of negative ions can be performed using a number of different techniques, and a common high precision approach is to perform tunable laser photodetachment spectroscopy (LPTS). One issue with this method is the possible presence of excited state detachment showing up as large background, causing a lower signal-to-noise ratio. A novel method recently developed at the DESIREE storage ring has solved this problem by selectively depleting all ions with electrons in excited states, leaving a pure ground state ion beam for LPTS measurements. This method has been demonstrated in the case of $^{16}$O$^-$, with the highest precision EA measurement to date\cite{Kristiansson2022High-precisionOxygen}. 

\begin{figure}
    \centering
    \includegraphics[trim={0cm 5cm 0cm 5cm},clip,width=0.4\textwidth]{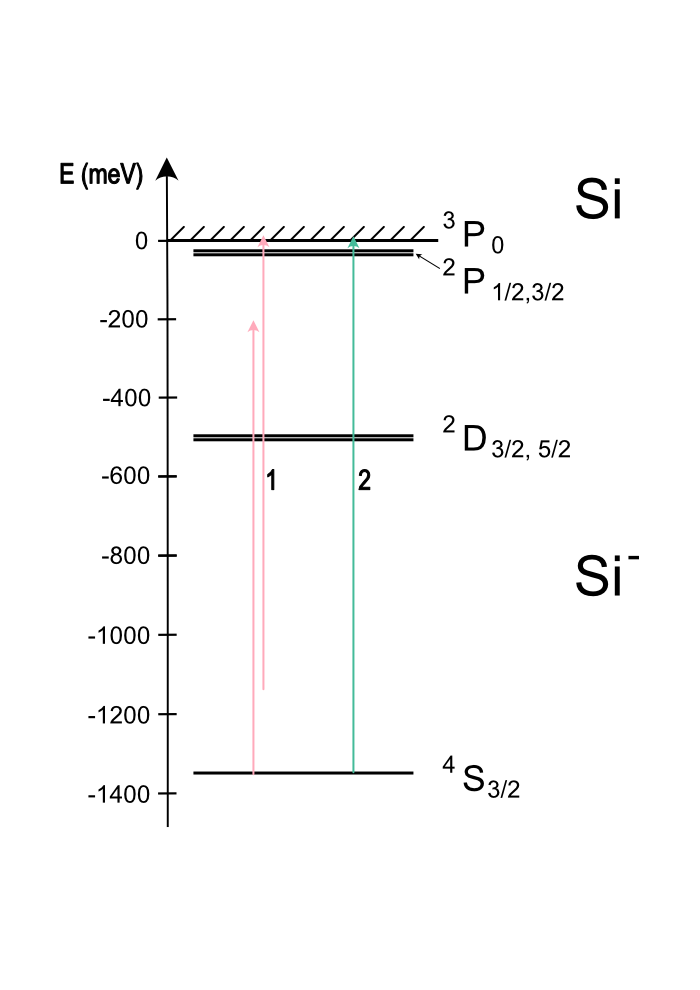}
    \caption{The energy levels of Si$^-$. Arrow set 1 represents the depletion of the excited states, whereas arrow 2 represents the photon energy to photodetach the ground state.}
    \label{fig:Si_levels}
\end{figure}

Silicon, with its energy level diagram shown in Fig. \ref{fig:Si_levels}, has three stable isotopes, $^{28}$Si, $^{29}$Si, and $^{30}$Si. The most abundant isotope is $^{28}$Si, for which an EA has been determined to 1.3895210(7) eV by Chaibi \textit{et al.} \cite{Chaibi2010EffectMicroscopy}. To the best of our knowledge, the other isotopes have not had their EAs experimentally determined prior to this work. Si$^-$ was recently investigated at the cryogenic storage ring (CSR) in Heidelberg \cite{Mull2021MetastableRing_2}. Here it was experimentally found that Si$^-$ has at least two long-lived excited states, and in the theoretical calculations, the four excited states were determined to have lifetimes ranging from  20 s for the $^2$P states up to several hours for the $^2$D states. This makes Si$^-$ an excellent candidate for the method used by Kristiansson \textit{et al.} in the case of $^{16}$O$^-$ \cite{Kristiansson2022High-precisionOxygen}.


In this paper we thus apply selective photodetachment to make a high resolution determination of the EAs for each of the three stable isotopes of Si$^-$. The resolution is sufficiently high to extract the isotope shift and even resolve the  HFS in the photodetachment threshold of  $^{29}$Si$^-$.

\section{\label{sec:experimentalmethods}Experimental methods}
The experiment was carried out in the cryogenic storage ring DESIREE, which is a double storage ring kept at a pressure of $2 \times 10^{-14}$ mbar at a temperature of 13 K. The low pressure is a prerequisite for the high signal-to-background ratio needed in the high precision measurement of the photodetachment threshold. To further decrease the background, the quantum-state populations were manipulated using selective photodetachment of all excited states in the negative ions. For $^{28}$Si$^-$, this was done using a Neodymium-doped yttrium lithium fluoride (Nd:YLF) laser with an output power of up to 5 W at a wavelength of 1053 nm. This laser was overlapped with the ion beam in the straight section of the symmetric storage ring, on the opposite side of the interaction region for the LPT spectroscopy. At the end of the straight section, a mirror reflects the laser beam back the same way it entered the chamber. In this way, the laser beam was directed back out of the chamber  and overlapped twice with the ion beam, allowing for faster depletion. 

\begin{figure}
    \centering
    \includegraphics[trim={2cm 4cm 0cm 2cm},clip,width=0.49\textwidth]{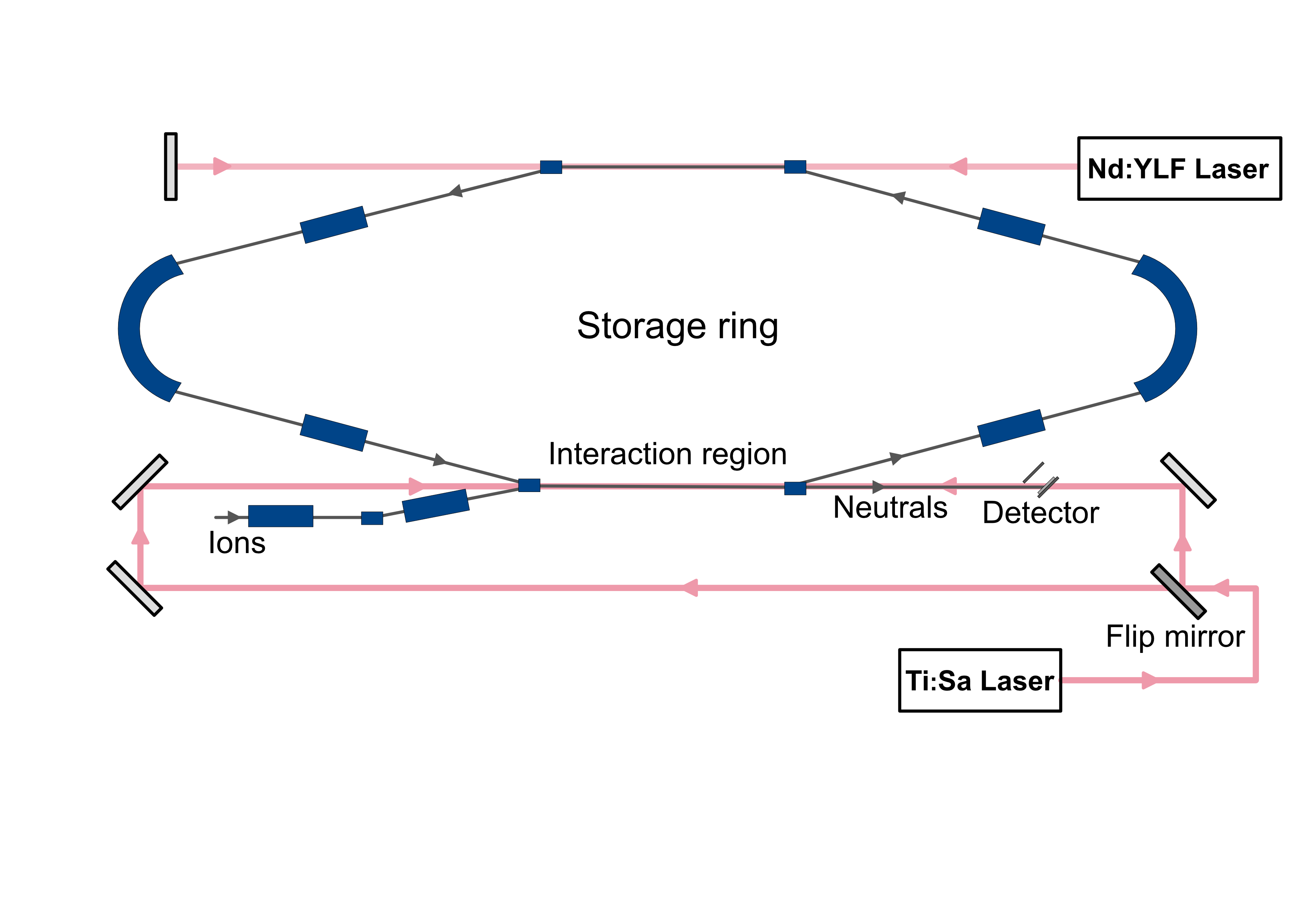}
    \caption{Experimental setup for the high precision EA measurements. The Nd:YLF laser was used for depletion of excited states and the Ti:Sa laser was used for photodetachment of the ground states in the Si isotopes. The experiment could be done in an alternating co- and counter-propagating geometry with the use of a mechanical flip mirror.}
    \label{fig:DESIREE_setup}
\end{figure}

For the heavier two isotopes there is a risk of silicon hydride contaminants in the ion beam. This cannot be detached by the 1053 nm wavelength of this laser, so in the case of $^{29}$Si$^-$ and $^{30}$Si$^-$ another laser was needed for detachment of SiH$^-$. Hence, a tunable SolsTiS narrow-linewidth titanium-sapphire (Ti:Sa) laser from M Squared \cite{M2} was used at a high power. Later during each cycle, the same laser was used for photodetchment of the ground state, but at a significantly reduced power. The full setup is shown in Figure \ref{fig:DESIREE_setup}.

This procedure leaves essentially all of the ions in the ground state, and the background is significantly reduced, allowing for high-precision measurements. 

The ions were injected into the storage ring with an energy of 25 keV and then circulated in the ring for 30 s, allowing the Nd:YLF laser to deplete the ions in excited states. In the case of $^{29,30}$Si$^-$, an additional laser was used for depletion, in the form of the Ti:Sa laser used for threshold scans. This means, for the first 30 seconds, this laser was set to a wavelength photodetaching atomic Si$^-$ ions in excited states as well as SiH$^-$. 

\begin{figure}
    \centering
    \includegraphics[trim={4cm 1cm 3cm 1cm},clip,width=0.49\textwidth]{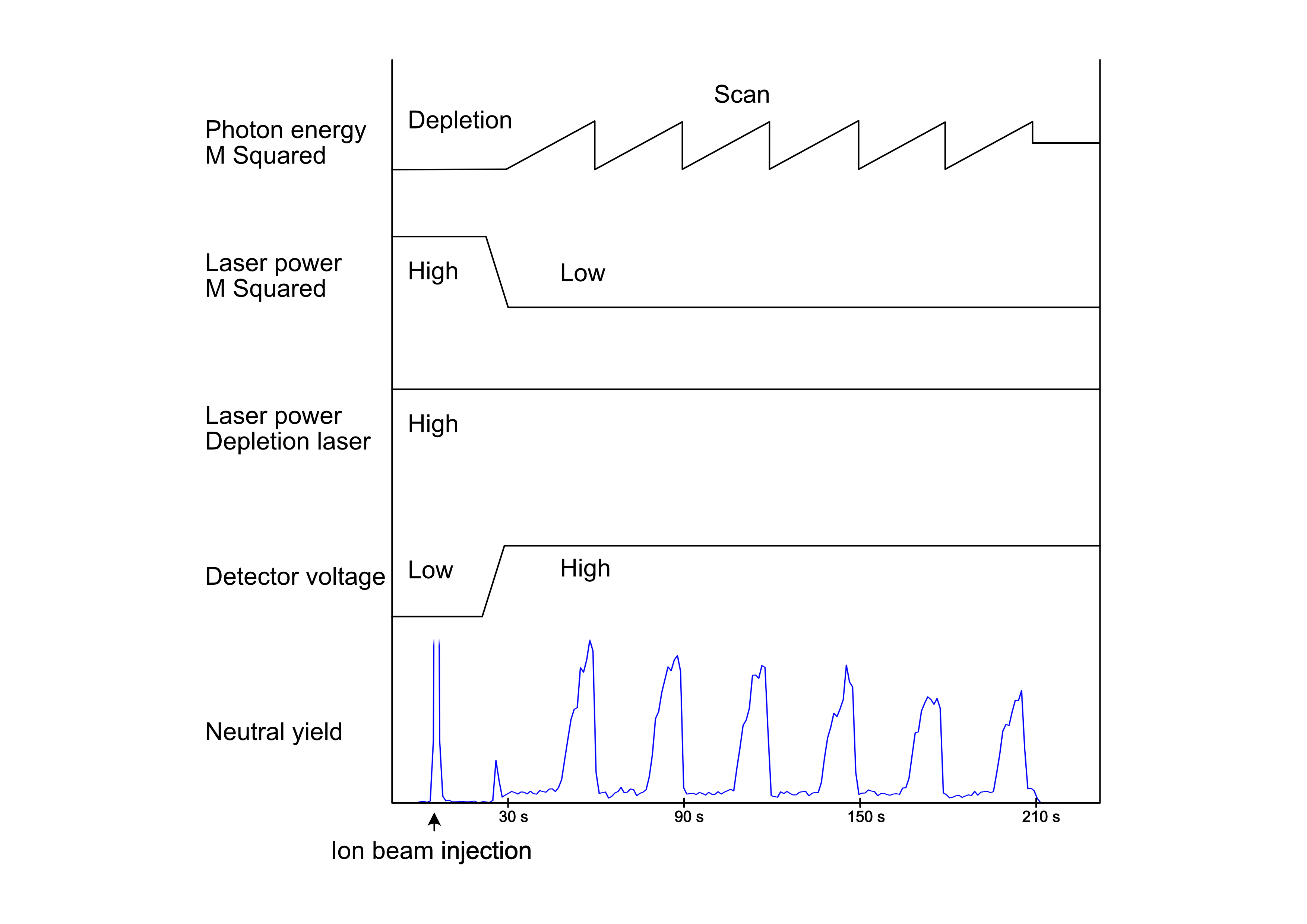}
    \caption{Timing of the experiment. During the first 30 seconds, the ions in excited states were depleted with high power lasers, whereupon six threshold scans were performed in total for each ion beam injection in the storage ring.}
    \label{fig:exp_timing}
\end{figure}

After this, the Ti:Sa laser was set to low power and the laser frequency was scanned over the EA region six times for every ion injection and the neutral yield was detected with a neutral particle detector. Subsequently, a flip mirror was inserted in the laser beam path according to Figure \ref{fig:DESIREE_setup}, and the whole measurement cycle was repeated with the Ti:Sa laser entering the DESIREE chamber from the opposite direction. In this way, the measurement was repeated in an alternating co- and counter-propagating geometry as discussed below. The timing for one measurement cycle is shown in Figure \ref{fig:exp_timing}.

The laser frequency was measured by a HighFinesse WS8-2 wavemeter with a precision of 10 MHz for a calibration frequency further away than about 170 MHz from the frequency in the region of interest. The calibration of the wavemeter was performed by a Toptica TA pro diode laser, which had been locked to the well-known quadrupole transition of 444.779,044,095,484,6(15) THz in $^5$S$_{1/2}$-$^4$D$_{5/2}$ in $^{88}$Sr$^+$ \cite{Margolis2005StrontiumUncertainty}. The frequency was periodically calibrated to make sure that the frequency readout was correct. 

\section{\label{sec:results}Results}
\subsection{Electron affinities and isotope shifts}
\label{sec:Results EA}
For electron affinity measurements, the EA was extracted from a fit made to the data using a convolution between the Wigner law and a Gaussian distribution. For Si$^-$, where $l = 0$ the Wigner law becomes 
\begin{equation}
    \sigma(E) = A + B \sqrt{E-E_\mathrm{EA}}\Theta(E,E_\mathrm{EA}),
\end{equation}
where $\Theta$ is the Heaviside step function, $E_\mathrm{EA}$ is the threshold energy and $A$ an $B$ are constants. The odd Si isotope has a HFS of the ground state, but for the EA measurement this was not resolved.

The experiment was performed in both a co- and counter-propagating setup, alternating directions every three cycles to account for any possible drift in ion velocity. Since the ions and laser either intersect in a co- or counter- propagating geometry, the resulting EA will be either red- or blue-shifted. To eliminate the Doppler shift, the geometric mean of the two results is used \cite{Juncar1985NewProposal}, which means we get 
\begin{equation}
\begin{aligned}
    E_{EA} &= \sqrt{\frac{1+v/c}{\sqrt{1-v^2/c^2}}E_{EA} \frac{1-v/c}{\sqrt{1-v^2/c^2}}E_{EA}} \\
    &= \sqrt{E_{EA}^{\uparrow\uparrow} E_{EA}^{\uparrow\downarrow}}, 
\end{aligned}
\end{equation}
where the co-propagating value for the EA is denoted by $E_{EA}^{\uparrow\uparrow}$ and the counter-propagating value is denoted by $E_{EA}^{\uparrow\downarrow}$.

\begin{figure}[p]
     \centering
    \begin{subfigure}{0.45\textwidth}
         \centering
         \includegraphics[width=\textwidth]{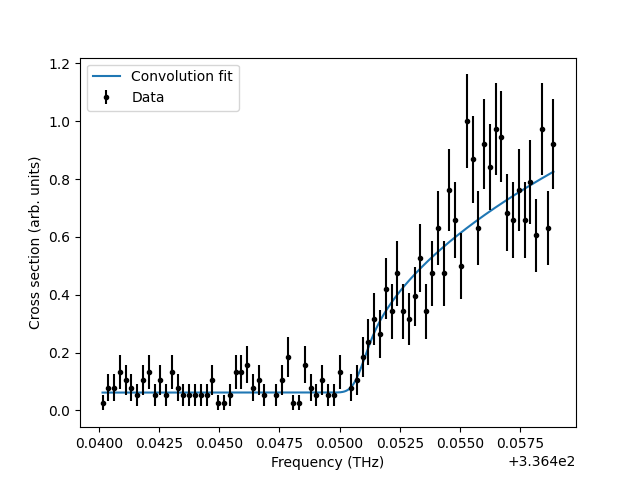}
         \caption{$^{28}$Si$^-$.}
         \label{fig:thr_28}
     \end{subfigure}
     \hfill
    \begin{subfigure}{0.45\textwidth}
         \centering
         \includegraphics[width=\textwidth]{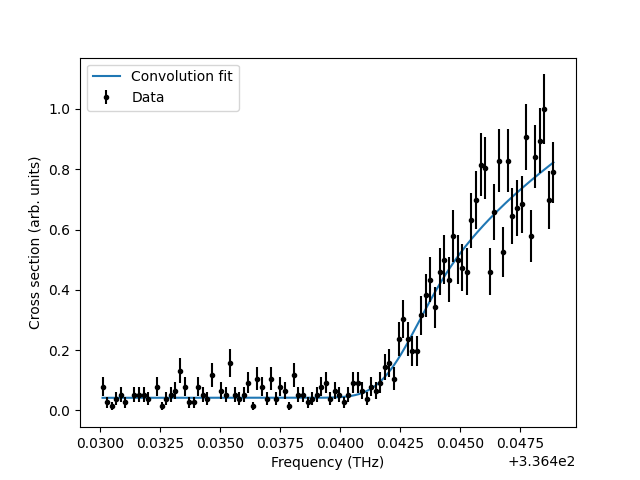}
         \caption{$^{29}$Si$^-$.}
         \label{fig:thr_29}
     \end{subfigure}
     \hfill
    \begin{subfigure}{0.45\textwidth}
         \centering
         \includegraphics[width=\textwidth]{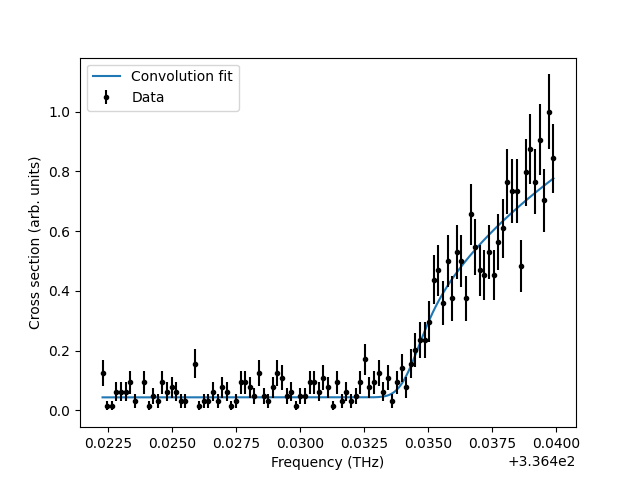}
         \caption{$^{30}$Si$^-$.}
         \label{fig:thr_30}
     \end{subfigure}
     \hfill
        \caption{Relative cross sections of the photodetachment thresholds in $^{28}$Si$^-$, $^{29}$Si$^-$, and $^{30}$Si$^-$. The blue solid line is a fit to the data of a convolution between the Wigner threshold law and a Gaussian distribution. All of these thresholds are from measurements performed in a co-propagating geometry.}
        \label{fig:LPT_28_30}
\end{figure}

The resulting electron affinities and isotope shifts are presented in Table \ref{tab:Si_EA_all}. Typical threshold scans for all three isotopes are shown in Figure \ref{fig:LPT_28_30}.  The blue lines represent fits of a convolution between the Wigner threshold law and a
Gaussian distribution.

\begin{table*}[]
    \centering
    \caption{Experimentally determined electron affinities and isotope shifts for the stable isotopes of Si$^-$.}
    \begin{threeparttable}
         \begin{tabular}{@{}lllll@{}}\toprule Isotope \hspace{0.2cm}& EA this work (eV)\hspace{0.2cm} &  EA previous work (eV) \hspace{0.2cm}& \multicolumn{2}{c}{Isotope Shift ($\upmu$eV)} \\\midrule 
        $^{28}$Si & 1.389 522 01(17) & 1.389 521 0(7) \cite{Chaibi2010EffectMicroscopy}  & $^{29-28}$Si \hspace{0.1cm} & -0.29(16) \\ 
        $^{29}$Si & 1.389 521 72(12)  & N/A & $^{30-29}$Si & -0.94(10)\\  
        $^{30}$Si & 1.389 520 78(12)  & N/A & $^{30-28}$Si & -1.23(16)\\  
        \bottomrule 
        \end{tabular} 
    \end{threeparttable} 
    \label{tab:Si_EA_all}
\end{table*}

The statistical errors of the EA measurements are $0.14 \upmu$eV, $0.064 \upmu$eV and $0.067 \upmu$eV  for $^{28}$Si, $^{29}$Si and $^{30}$Si, respectively. In addition to this, systematic errors are added to the final uncertainties in the EAs. The uncertainty in the measurement of the wavemeter is expected to be 10 MHz ($0.041 \upmu$ eV), and the uncertainty in the photodetachment signal due to laser power fluctuations is set to 15 MHz ($0.062 \upmu$eV). The photodetachment of the stored ions will affect the velocity distribution with time depending on the applied laser power, and even though the laser power was reduced in order to eliminate this effect, a conservative value of 15 MHz ($0.062 \upmu$eV) is added to the uncertainty. This results in the total uncertainties of $0.17 \upmu$eV, $0.12 \upmu$eV and $0.12 \upmu$eV, for $^{28}$Si, $^{29}$Si and $^{30}$Si, respectively. This gives us the resulting EAs presented in Table \ref{tab:Si_EA_all}. For the isotope shifts, only the statistical errors are included, as all other uncertainties cancel to first order.

\subsection{Hyperfine structure of $^{29}$Si$^-$} 

\begin{figure}[ht]
    \centering
    \includegraphics[width=0.45\textwidth]{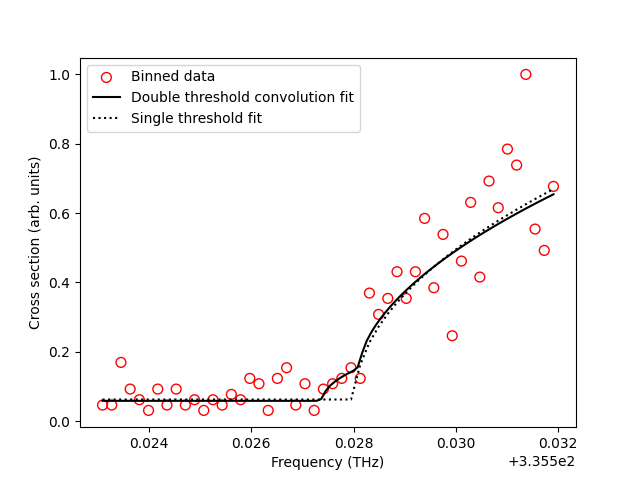}
    \caption{Hyperfine structure of $^{29}$Si$^-$. The solid line is a fit to the Wigner  double threshold law convoluted with a Gaussian distribution. The dashed line is a fit to a single threshold for comparison. This measurement was performed in a counter-propagating geometry.}
\label{fig:29_noconv_conv}
\end{figure}

In the data shown in Fig. \ref{fig:LPT_28_30} we observed that the width obtained in the fitting is larger for the $^{29}$Si$^-$ as compared with $^{28}$Si$^-$ and $^{30}$Si$^-$. This is likely due to  HFS in the odd isotope. Figure \ref{fig:29_noconv_conv} shows a high resolution scan over the threshold of $^{29}$Si$^-$ made to investigate this range further. Here a small scan range, a smaller step and longer integration time was used as compared with the data shown in Figure \ref{fig:LPT_28_30}. With these experimental parameters we resolve a small HFS in the photodetachment threshold. It should here be pointed out that  $^{29}$Si$^-$ is particularly suitable to observe the HFS. 
$^{29}$Si$^-$ has a nuclear spin I=1/2, and the ground state of the negative ion has an total angular moment of  J=3/2 giving two HFS levels $F = 1,2$. The groundstate of neutral silicon, on the other hand, has J=0 and hence no HFS. This means that there will only be two overlapping thresholds corresponding to opening of the F=1 and F= 2 HFS components of the groundstate of the negative ion. The data was analyzed by making a fit  to a double Wigner threshold law
\begin{equation}
\begin{aligned}
    \sigma(E) = A + B \left[(2F+1) \sqrt{E-E_{\mathrm{EA}}}\Theta(E,E_{\mathrm{EA}}) \right.\\
    \left. + (2F'+1) \sqrt{E-E_{\mathrm{EA+hfs}}}\Theta(E,E_{\mathrm{EA+hfs}}) \right],
\end{aligned}
\label{eq:Wigner_hfs}
\end{equation}
where $A$ and $B$ are constants, the total angular momentum $F = 1$ and $F' = 2$, $E_\mathrm{EA}$ is the electron affinity of $^{29}$Si and $E_\mathrm{EA+hfs}$ is the energy of the second energy threshold. $\Theta$ is again the Heaviside step function. The relative weight of the two channels is assumed to be given by the degeneracy for this very small excitation energy.

We observed a minute drift of the threshold due to small fluctuations of the ion beam energy, and hence the ion velocity. Therefore, a fit was made to each individual scan. This yielded in a total HFS  splitting of the ground state of $^{29}$Si$^-$ {$1.8(4)\upmu$eV}. The statistical uncertainty was {$0.2\upmu$eV}. The systematic uncertainties discussed in section \ref{sec:Results EA} will affect the two thresholds equal and will hence cancel in the isotopes shifts. However, we do see a small effect that the measured HFS was slightly larger for thresholds measured at the end of each measurement cycle.  As a conservative estimate of this effect we add {$0.2\upmu$eV} to the uncertainty giving the total uncertainty of {$0.4\upmu$eV}. 

\section{\label{sec:discussion}Discussion}
The method used in this experiment, where a laser is used at high power to deplete the ions in excited states while leaving the ground state ions untouched in the storage ring and then selectively photodetach ions in the ground state, was first used in electron affinity measurements at DESIREE by Kristiansson \textit{et al.} for $^{16}$O$^-$ \cite{Kristiansson2022High-precisionOxygen}. This resulted in the highest precision EA to date, and was followed by an EA measurement of $^{18}$O$^-$ and the corresponding isotope shift \cite{Kristiansson2022PrecisionIons}. In this work, the same method was used, improved by the addition of a second high power laser to avoid changing the wavelength and power of the laser used for the LPT spectroscopy. 
{However, due to the contamination of SiH$^-$ for the $^{29}$Si$^-$ and $^{30}$Si$^-$ isotopes, the Ti:Sa laser had to be used for both depletion of excited states and LPTS of the ground states. Also, the high power Nd:YLF laser unfortunately minutely heated up the DESIREE chamber when applied at maximum power, so this had to be attenuated to a power of approximately 2 W. This was likely due to laser light heating up some components inside the vacuum chamber.} 


The EA of three isotopes of Si$^-$ have been measured, and in the case of $^{29}$Si and $^{30}$Si, there are no previous results for comparison. The new measured value of 1.389 522 01(17) eV for $^{28}$Si$^-$, which is a fourfold improvement in the accuracy as compared with the previous value of 1.389 521 0(7) eV measured by Chaibi \textit{et al.} \cite{Chaibi2010EffectMicroscopy}. The two values agree within 
 with $2 \sigma$.   



In the EA measurement of $^{29}$Si$^-$, the hyperfine structure could not be resolved, which means that the EA was determined using a regular one-threshold Wigner law. This value is thus not corrected for the hyperfine splitting. To resolve the hyperfine structure, the wavelength scan region was greatly reduced, and a lot of statistics were gathered. $^{29}$Si$^-$ is the first measured hyperfine structure observed in a threhold measurement.

\section{Conclusions}
The EAs and isotope shift for the three stable isotopes of Si$^-$ have been measured at DESIREE, using a new experimental technique  \cite{Kristiansson2022High-precisionOxygen},where laser-manipulation of quantum-state populations was used to prepare a ground state beam of negative ions. In this way, high precision values for the EAs have been achieved, and the $\upmu$eV isotope shifts in the EA have been successfully measured. In the odd isotope, the hyperfine structure splitting could also be measured due the extremely low background signal and the narrow-linewidth, high-precision laser system.

To the best of our knowledge, only two prior experimental observations of hyperfine structure in negative ions have been made. These applied rather different techniques \cite{Trainham1989Measurement33S-,Fischer2010FirstAnion}, making this experiment the first of its kind. This high precision method of measuring the EA can be applied to other odd isotopes in order to determine the hyperfine splittings and to enrich our understanding of negative ion structure. Additionally, such high precision isotope shift measurements, combined with state of the art specific mass shift calculations, allow the extraction of filed shift values. A reliable method for determining the field shifts in atomic systems would be of great value to nuclear physics.

Apart from measuring energy levels to a high precision, this method of state manipulation can be used to investigate i.e. the mutual neutralization (MN) process. Both the positive and the negative ions can be selected to be in the ground states before the mutual neutralization takes place. In this way, it is possible to compare manipulated quantum level MN and the same process with ions in any possible energy state. 

\section{Acknowledgements}
{DH, HZ, HC and HS acknowledge support from the Swedish Research Council under contracts 2020-03505, 2020-03437, 2023-03833 and 2022-02822. This work was performed at the Swedish National Research Infrastructure, DESIREE (Swedish Research Council Contract No. 2017-00621 and No. 2021-00155). We want to thank the staff at DESIREE for their support during the beamtime. The Trapped Ion Quantum Technologies
group at Stockholm University is acknowledge for providing the wavemeter calibration signal. This work was supported in part  (NDG)  by U.S. NSF Grant
No. PHY-2110444. 
 }

\appendix

\bibliographystyle{unsrt}
\bibliography{references,references2}

\end{document}